\documentclass[3p, 12pt]{elsarticle}

\usepackage{graphicx}
\usepackage[latin2]{inputenc}
\usepackage{hyperref}
\hypersetup{
      colorlinks=true,
    linkcolor=blue   ,
    citecolor=blue,
    filecolor=blue,
    urlcolor=blue
    }
 \biboptions{sort,compress}

\begin{document}

\title{Fabrication of ballistic suspended graphene with local-gating}

\author[ba]{Romain Maurand}
\author[ba]{Peter Rickhaus}
\author[ba]{P\'{e}ter Makk\corref{cor1}\fnref{fn1}}
\ead{peter.makk@unibas.ch}
\cortext[cor1]{Corresponding author}
\author[ba]{Samuel Hess}
\author[bp]{Endre T\'ov\'ari}
\author[ba]{Clevin Handschin}
\author[ba]{Markus Weiss}
\author[ba]{Christian Sch\"onenberger}

\address[ba]{Department of Physics, University of Basel, Klingelbergstrasse 82, CH-4056 Basel, Switzerland}
\address[bp]{Department of Physics, Budapest University of Technology and Economics and Condensed Matter Research Group of the Hungarian Academy of Sciences, Budafoki ut 8, 1111 Budapest, Hungary
}

\date{\today}

\begin{abstract}
Herein we discuss the fabrication of ballistic suspended graphene nanostructures supplemented with local gating. Using in-situ current annealing, we show that exceptional high mobilities can be obtained in these devices. A detailed description is given of the fabrication of bottom and different top-gate structures, which enable the realization of complex graphene structures. We have studied the basic building block, the p-n junction in detail, where a striking oscillating pattern was observed, which can be traced back to Fabry-Perot oscillations that are localized in the electronic cavities formed by the local gates. Finally we show some examples how the method can be extended to incorporate multi-terminal junctions or shaped graphene. The structures discussed here enable the access to electron-optics experiments in ballistic graphene.
\end{abstract}

\maketitle

\section{Introduction}

Several recent experiments, like transverse  magnetic focusing \cite{focusing13}, the observation of superlattice effects \cite{Ponomarenko13,dean13} complex ground-state structure of bilayer graphene \cite{PhysRevLett.108.076602, Bao08062012}, fractional quantum Hall effect \cite{xu09, Ki13} and large amount of theoretical proposals \cite{PhysRevLett.97.067007,Cheianov02032007}  rely on the realization of ballistic graphene devices with dedicated local gating. Nowadays, a large part of the graphene community is using stacks of graphene/hexagonal boron nitride (h-BN) \cite{doi:10.1021/nl2005115,Dean2010,Wang01112013,xue2012,Yu11022013,JJAP.52.110105, engels13} to obtain high-mobility devices.  However, several recent experiments show that the level of quality of freely suspended devices obtained by the method first proposed by Tombros et. al. \cite{tombros11} is equivalent or even higher than graphene encapsulated in h-BN \cite{grushina,rickhaus13}. In this paper we report on our experience in fabricating freely suspended graphene samples following this method. We first discuss in detail the key elements of fabrication and the modification to the original method which led us to the production of ultra-high quality devices with an excellent yield.  Quality of the samples will be discussed based on transport experiments revealing ballistic interferences as well as quantum Hall plateaus at very low magnetic field. Afterwards we will show our implementation of local gates, bottom and top gates, for these suspended devices and demonstrate that a ballistic p-n junction can be realized. Finally we discuss the feasibility of more complex graphene devices like multi-terminal samples or Aharonov-Bohm rings.

\section{Detailed methods}

\begin{figure}[!htb]
\begin{center}
\includegraphics[width=\columnwidth]{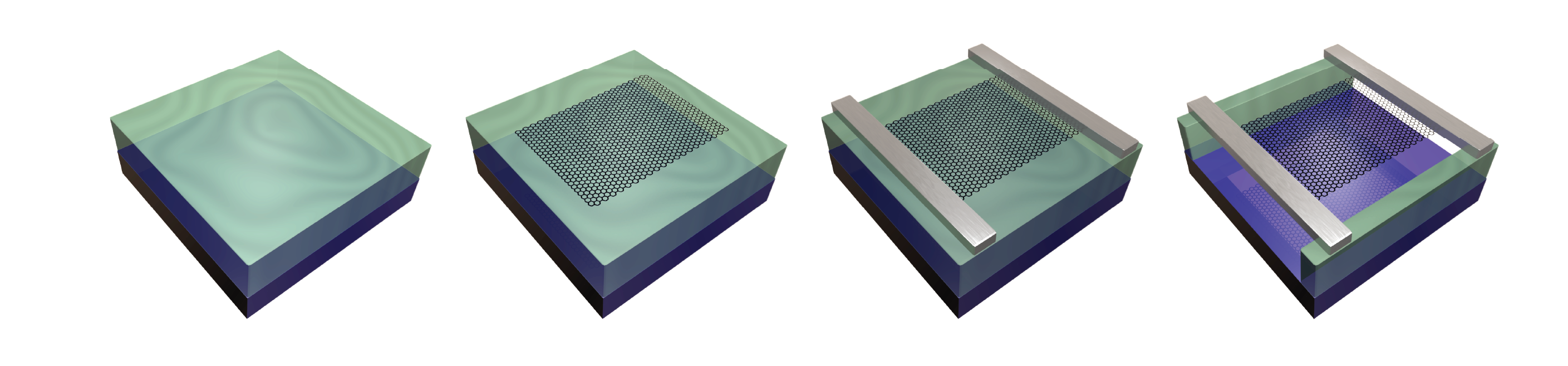}
\end{center}
\caption{\emph{Fabrication of suspended graphene following the method introduced by Tombros et al. \cite{tombros11}. The SiO$_2$ is colored blue, whereas the LOR is shown in green. The final image shows the device suspended together with the contacts. }} \label{Fig1}
\end{figure}

The main idea of the fabrication method proposed by Tombros and coworkers \cite{tombrosNatPhys,tombros11} is to use a layer of lift-off resist (LOR) \footnote{LOR 5A, MicroChem Corp.} as a sacrificial layer to realize freely suspended graphene devices. Figure~\ref{Fig1} summarizes this method. We start the fabrication by spin coating a heavily doped Si/SiO$_2$ wafer with LOR. Compared to Tombros and coworkers we reduced the thickness of the LOR from 1.15 $\mu$m to 600$\,$nm in order to increase the back-gate efficiency to access higher charge carrier densities. Even with this reduced thickness, the suspension remains possible without a critical point drying system. Afterwards graphene from natural graphite is exfoliated with Nitto tape\footnote{SPV 224P, Nitto Europe NV} directly onto the LOR surface.

No particular care is required for the identification of graphene by optical contrast on the wafer/resist stack, since we use dark field and Nomarski differential interference contrast (NDIC) microscopy to locate single layer graphene flakes. NDIC microscopy separates the source light into two orthogonally polarized parts, which are recombined after reflection on the device. The contrast given by the interference of the two polarized parts reveals extremely well the edges of graphene flakes \cite{axon86}. Also, the microscope is equipped with a camera system, which allows digital contrast enhancement of the live microscope image, providing an easy determination of the number of graphene layers. Once a suitable graphene flake has been spotted, the structures are realized by e-beam lithography with a 300$\,$nm thick 950K PMMA layer at 20$\,$keV electron acceleration voltage with a dose of 200 $\mu$C/cm$^2$. As for Tombros et al., PMMA development is done with xylene at room temperature.  After development, 50$\,$nm of palladium is thermally evaporated in a home-made UHV chamber with the device cooled to -30$^\circ$C. Lift-off is done with xylene at 80$^\circ$C, and the sample is rinsed with hexane. We note here that micro-bonding of these samples is hard, since the bonding pads are on LOR. This problem can be circumvented if the LOR is removed at the bonding pads before the contact fabrication.  To suspend graphene, LOR is exposed using e-beam lithography at 20 keV with a dose of 1100 $\mu$C/cm$^2$. Since the resist layer is only 600$\,$nm thick and due to back-reflected electrons, the LOR is exposed below the contact. Finally after development of LOR with ethyl-lactate both graphene and metallic contacts are suspended. This is different from the original method where the contacts were supported by LOR pillars at the end of the suspension. We have found that what we call "full-suspension" is crucial for an efficient current annealing.

\section{Current annealing and transport experiments}

\begin{figure}[!htb]
\begin{center}
\includegraphics[width=\columnwidth]{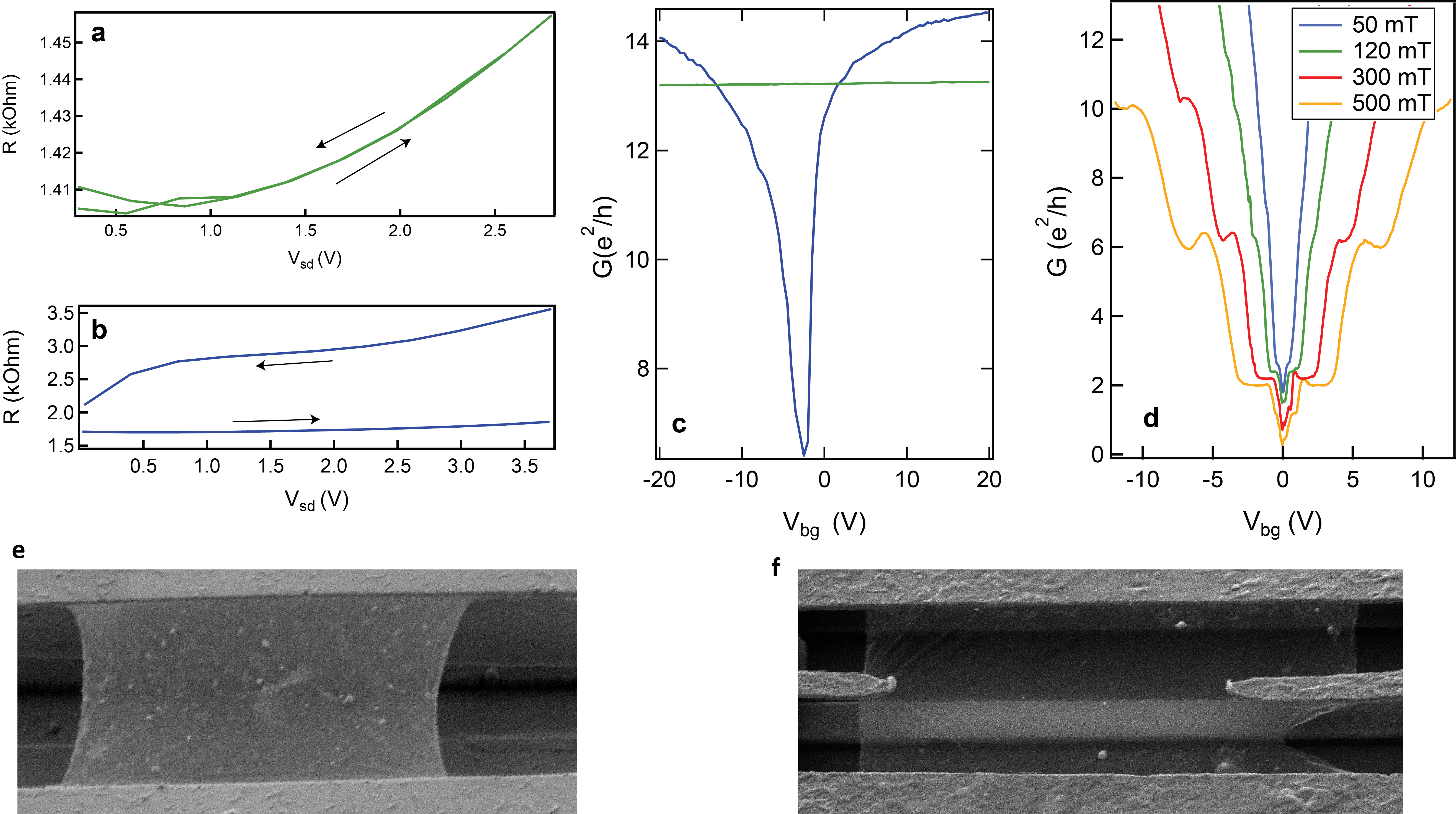}
\end{center}
\caption{\emph{Typical current annealing resistance-voltage traces for the first (a) and last annealing step (b). c) Graphene conductance dependence on back gate before current annealing (green), and after the last annealing step (blue). d) Quantum Hall effect measurement at low fields at T=1.5$\,$K. The first plateau is pronounced already at 50$\,$mT. The individual curves are shifted by 0.2$\,e^2/h$ for clarity. e) SEM image of a multi-terminal graphene device before current annealing. The graphene is covered with resist residues. f) Another multi-terminal device after current annealing, showing reduced number of residues.}} \label{Fig2}
\end{figure}

Electronic transport experiments have been done either in a 4He cryostat with a variable temperature insert at low gas pressures $<$1$\,$mbar or in a dilution refrigerator depending on the base temperature needed. In both cases current annealing was performed in order to clean graphene. The annealing was done at 4$\,$K, where we have a better control over the annealing process compared to room temperature.  Just after fabrication, graphene devices only show a very weak field effect indicating a strong doping by resist residues from LOR and PMMA, as shown by the green curve in Fig.~\ref{Fig2}c).  Our initial studies on supported samples show current annealing yield below 20$\%$, however,  the "full-suspension" fabrication allows obtaining ultra-clean graphene devices with a current annealing yield close to 100$\%$, and these devices are at the state of the art concerning residual doping and ballistic behaviour.

For the annealing we apply a voltage to the device in series with a resistor of 1$\,$kOhm, while we record the resistance of the device. During an annealing cycle we slowly ramp-up the voltage to a maximum value and ramp it down to zero. The full-annealing procedure finally consists of repeating this cycle while increasing the maximum voltage applied.  The first cycle is done with V$_{max} = 0.5\,$V. If a resistance difference between the increasing and the decreasing voltage sweep is found, the gate dependence is recorded.

In Figure~\ref{Fig2}a) and b) current annealing curves for a "fully-suspended" device with the corresponding gate traces are shown.  Before annealing, the graphene device displays only a very small field effect as shown by the green curve in Figure~\ref{Fig2}c). In Figure~\ref{Fig2}a) we plot the resistance of the device during an annealing cycle with V$_{max} = 3\,$V.  This annealing curve displays no hysteresis in the recorded resistance; the graphene quality remains similar and still shows weak gate dependence.  In Figure~\ref{Fig2}b), where the last annealing cycle is shown, V$_{max}$ is increased to 3.75$\,$V,  and a large resistance jump is visible leading to a hysteretic annealing curve.  The gate dependence measured after this step is shown in blue in Figure~\ref{Fig2}c) and displays a sharp conductance minimum corresponding to the charge neutrality point (CNP or Dirac point).  We have noticed that the current densities needed to completely clean suspended devices, which for 2-terminal devices are in the order of $\sim350 \pm 120\,\mu$A$/\mu$m per $\mu$m flake width, are smaller than current densities reported for devices with supported contacts \cite{Bolotin2008351}. During the current annealing the current heats up graphene, but the contacts remain colder where residues can condensate. When the contacts are also suspended, the cooling power of the contacts from the substrate is reduced, and the temperature of the contacts also increases during the annealing. This ensures that the graphene is also cleaned around the contacts, not only at the middle of the flake. The CNP is almost at zero gate voltage revealing a weak residual doping of maximum n$\sim$10$^9\,$cm$^{-2}$. Fully suspended devices show mobilities $\mu=\sigma/ne$, that are always higher than 100.000$\,$cm$^2$/Vs, where $n$ is determined by a simple capacitance model.  Moreover conductance oscillations visible on the blue curve of Fig.~\ref{Fig2}c) are signatures of Fabry-P\'{e}rot interferences over the full length of graphene, i.e. the electrons are bouncing back and forth between contacts. In  Figure~\ref{Fig2}e) an SEM image of a multi-terminal device is shown before current annealing, where resist residues from prior fabrication procedure are clearly visible. Another sample after current annealing is shown in Figure~\ref{Fig2}f), where the amount of residues is substantially lower than before annealing.
The low potential disorder of a device can also be judged from the lowest magnetic fields, at which the quantum Hall plateaus of conductance become visible. Figure~\ref{Fig2} shows two-terminal quantum Hall measurements for different magnetic fields. At $200\,$mT clear conductance plateaus at 2, 6 and 10 $e^2/h$ are distinguishable both for electron and hole doping (the aspect ratio of the device is 1:1, 2 by 2$\,\mu$m, therefore, the quantum Hall resistance plateaus are clearly visible in the two terminal data \cite{PhysRevB.80.045408}). However, at even lower field, at 120$\,$mT the plateaus are clearly seen, and the lowest plateau can even be seen at $50\,$mT. The "full-suspension" fabrication allows obtaining ultra-clean graphene devices with a current annealing yield close to 100$\%$, and these devices are at the state of the art concerning residual doping and ballistic behavior.

\section{Implementation of local gates}

\begin{figure}[!htb]
\begin{center}
\includegraphics[width=\columnwidth]{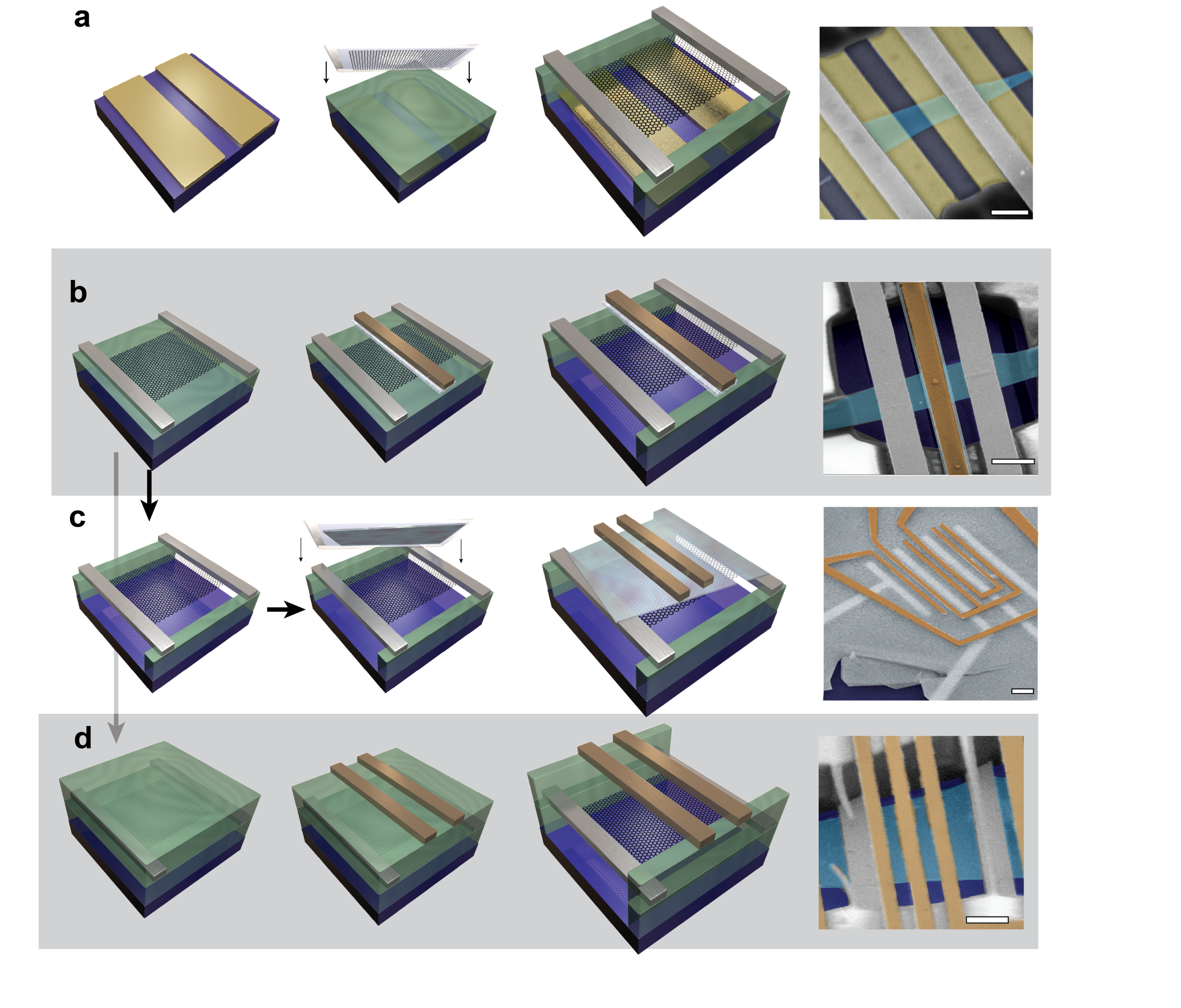}
\end{center}
\caption{\emph{Fabrication of gated structures, where the final images in each row are the false colored SEM images of the finished devices. The scalebar corresponds to 1$\,\mu$m. a) Fabrication of suspended devices with bottom gate structures. After fabrication of the gates, the graphene is transferred onto LOR. b)-d) Fabrication of top gates. b) The top gates are separated from graphene by a thin layer of MgO. c)  An additional layer of hBN is transferred onto graphene on which the gates are patterned. d) The graphene is covered with an additional layer of LOR, which is used to suspend the gates. In the SEM images the bottom and top gates are colored yellow and orange, respectively, and the contacts and graphene are shown by grey and light-blue, respectively.  }} \label{Fig3}
\end{figure}

In order to further explore the promising physics of Dirac fermions by electronic transport experiments \cite{Cheianov02032007}, it is essential to be able to tune the charge carrier density locally on a graphene device. This has already been done on supported samples both with bottom and top gates \cite{Eroms_Slattice,doi:10.1021/nl4006029,PhysRevLett.98.236803,Williams03082007}. Concerning suspended graphene, implementation of local top and bottom gates remains challenging, even if some realizations have been presented in the community \cite{allen12,velasco12,Weitz05112010}. Figure~\ref{Fig3} describes the implementation of local gates.
One direction to realize local gating is to pre-pattern bottom gates before the spin-coating of LOR, as shown in Figure~\ref{Fig3}a). The difficulty remains in depositing the desired graphene flake on top of these gates. To do so, a dry transfer technique has been chosen \cite{Dean2010}: graphene is exfoliated onto a separate wafer of Si/Si0$_2$ with a stack of PVA/PMMA of 50$\,$nm/300$\,$nm. Once again optical localization of single layer graphene flakes is done by NDIC microscopy, and no specific resists stack is needed for this. Afterward this wafer piece is put at the surface of DI-water, and the PVA layer is slowly dissolved. The floating PMMA layer (with the graphene flake on top) is carefully fished out with a glass slide containing a metallic rim (which will support the PMMA layer) and a hole on the glass slide, where the water can flow out \cite{Dean2010}. Then the glass slide is transferred to a modified MJB3 mask aligner to realize the alignment between the graphene flake and the pre-designed bottom gates coated with LOR.  The chip containing the bottom gates is heated to 120$^\circ$C for the transfer to avoid any water at the surface of LOR. This is shown in the second sketch of Fig.~\ref{Fig3}a). Afterwards, the stack is heated to 150$^\circ$C where the PMMA is detached from the glass slide, and the graphene is now transferred to LOR. The PMMA is removed by hot xylene and cleaned by hexane. Once graphene is transferred, the sample is finished following the fabrication method described previously in Figure~\ref{Fig1}.  The last sketches in Figure~\ref{Fig3}a) show the finished devices and a false-colored scanning electron micrograph of graphene suspended above the bottom gates.

Another possibility for gating the graphene flake is to fabricate top gates. However, the realization of top gates for a suspended device is not straightforward.  The most natural way is to grow a dielectric between graphene and the gate electrodes. For this we have evaporated 5-10$\,$nm of MgO below the top gates. The design and a false colored SEM image of the device are shown in Fig.~\ref{Fig3}b), where the top-gate is colored orange. We have managed to current anneal the device, with a Dirac point visible in G(V$_{bg}$) similar to Fig.~\ref{Fig2}c). However, we have found that the yield of current annealing is decreased, because of the middle top gate which acts as a heat sink, and the top gates seem to introduce noise possibly because of the trap states in MgO.
Instead of using an evaporated oxide for the dielectric, a layered material, e.g. hexagonal boron nitride (h-BN) can act as a spacer between graphene and the top-gates, and also supports the contacts.  It was shown already, that h-BN can be used for top gates in substrate supported samples, and stable gating can be realized \cite{Dean20121275}.  Therefore we have transferred h-BN flakes on top of the contacts of an already suspended device, as shown in Fig.~\ref{Fig3}c). Finally the top gates were fabricated on the h-BN.  A false-colored SEM image of a device is displayed in Fig.~\ref{Fig3}c), where the boron-nitride flake covering the whole structure is shown in light blue.  We have found however, that the yield of current annealing, similarly to the oxide defined top gates decreased significantly. Probably, the encapsulation of the device inhibited the evaporation of the contaminants from the device, and the hBN also acted as a heat sink during the annealing. Although we have managed to clean some of the devices, the decrease of the yield suggested looking for a different technique.
The problems of the previous two approaches can be circumvented if the top gates are also suspended above graphene. For this, after the fabrication of metallic contacts to graphene, another layer of LOR resist of thickness 500-600$\,$nm is spin-coated on top of the device, as shown in Figure~\ref{Fig3}d). Between the two layers of LOR a thin (80$\,$nm) layer of PMMA is added, since without this protection layer during the spin-coating of the second layer of LOR the contacts can be deformed, and no electrical contact between the Pd contacts and graphene can be measured. Top gates are designed by e-beam lithography using the same parameters as for the metallic contacts. Finally the whole device (containing graphene, contacts to graphene and the top gates) is suspended by exposing the suspension mask using e-beam lithography with a dose of 1100$\,\mu$C/cm$^2$ and by development in ethyl-lactate.

An SEM micrograph in Fig.~\ref{Fig3}d) shows such a device after fabrication, with the top gates displayed in orange, as before. If the gates cover a substantial part of the graphene flake, the suspension cannot be done as the final lithography step. In this case the suspension mask is exposed before the fabrication of the top gates, but the mask is not developed. To finalize the device, after the lift-off for the top gates, the device is suspended by developing the suspension mask in ethyl-lactate. The versatility of the fabrication method allows the combination of top gates and bottom gates to realize complex gated structures.  For example a local perpendicular electrical field to the graphene flake, created by local bottom and top-gates can be used to study and exploit the different ground states of ultra-clean bilayer graphene \cite{Bao08062012,PhysRevLett.108.076602}.

\section{p-n junctions}

\begin{figure}[!htb]
\begin{center}
\includegraphics[width=\columnwidth]{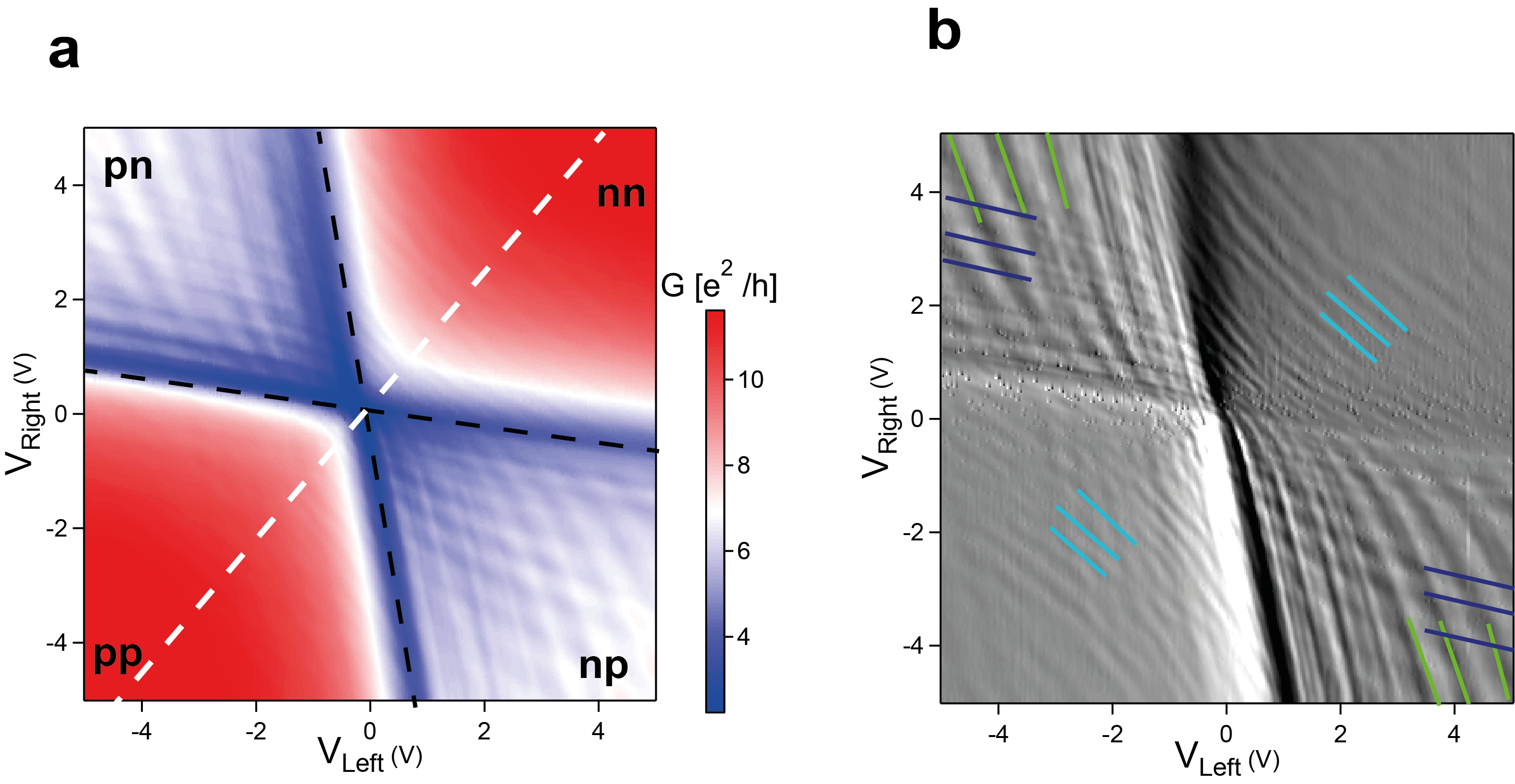}
\end{center}
\caption{\emph{a) Conductance map of a p-n junction as a function of two bottom gate voltages. In the p-n region the conductance is lower (blue) than in the unipolar regime (red).  Interference patterns can be seen in the hetero- and unipolar regions. b) Derivative of map a) with respect to V$_{right}$ showing pronounced Fabry-P\'{e}rot resonances. The different lines are marked by different colors: blue and green correspond to resonances in the right and left cavities in the bipolar regime, respectively, whereas the light blue resonances mark the FP resonances in the whole cavity in the unipolar case.}} \label{Fig4}
\end{figure}

Figure~\ref{Fig4} shows the differential conductance of a graphene flake suspended over two bottom gates, as a function of the applied bottom gate voltages. The design of such structure is shown in Fig.~\ref{Fig3}a): one bottom gate is located below the left part of graphene and the other one below the right part, and the charge carrier density in the left and right side are tuned by the corresponding gate voltages. The plot can be divided into four quadrants, also marked by the black dashed line, where the quadrants correspond to different configuration of carrier type in the left and right part of graphene \cite{rickhaus13}. The almost horizontal and vertical dashed lines mark the CNP of the right and left part, respectively. When both gate voltages are negative, the whole suspended graphene is p-doped, and similarly, for positive voltages on both electrodes it is electron doped. These two regimes are called unipolar. When the polarity of the two gates is opposite, doping on the left and on the right side of graphene is different and a p-n interface is created in the middle of the device. This is called the bipolar regime, and as a consequence of the potential step (p-n interface) the conductance in the bipolar regime is lower than in the unipolar. This particular geometry allows the precise study of ultra-clean p-n junction.

Fig.~\ref{Fig4}b shows the derivative of conductance with respect to the right bottom gate voltage. The map shows fine lines in the unipolar regime, which correspond to interference of electrons that are reflected from the contacts, thus forming an electronic Fabry-Perot cavity over the whole device length. These resonances are marked by light blue lines. The appearance of resonances shows, that the electronic transport is ballistic over the whole cavity. Further lines appear in the bipolar regime running parallel to the Dirac-lines separating the quadrants, which originate from resonances in the left and the right cavity. These lines are also marked in Fig.~\ref{Fig4}b) with dark blue and green, respectively. We have given a detailed analysis of the Fabry-Perot resonances in Ref. \cite{rickhaus13}, for a related study see Ref. \cite{grushina}.

\section{Further possibilities of the method}

\begin{figure}[!htb]
\begin{center}
\includegraphics[width=\columnwidth]{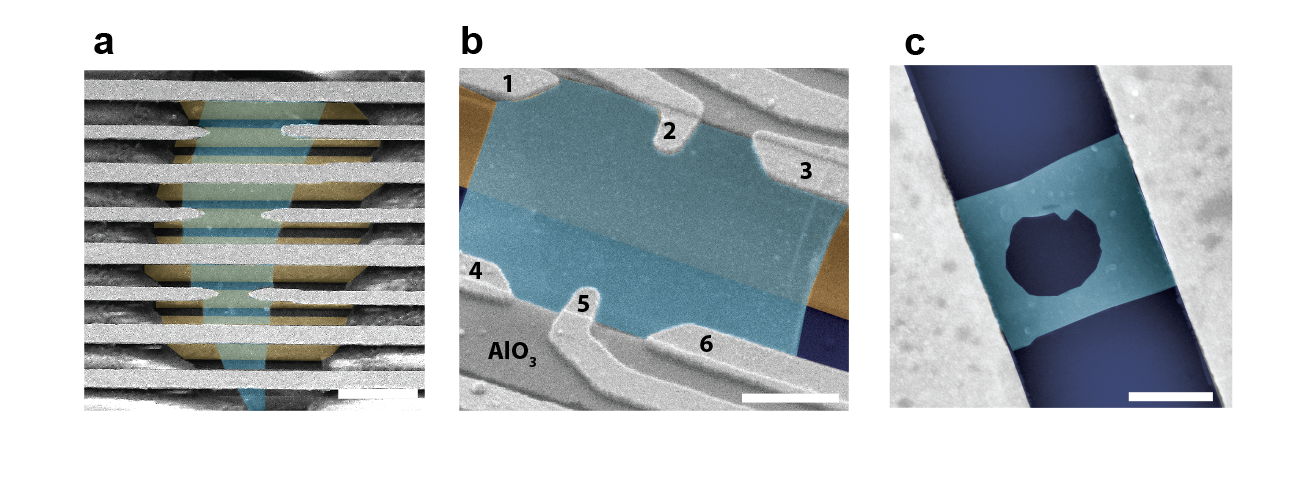}
\end{center}
\caption{\emph{False colored SEM images of more complicated structures. The scalebar corresponds to 1$\,\mu$m. a) Multi-terminal device with side contacts and bottom gates. b) Multi terminal device with finger electrodes and bottom gates.  c) Suspended Aharonov-Bohm ring, defined by plasma etching.}} \label{Fig5}
\end{figure}

Finally to emphasize the great versatility of this fabrication method Figure~\ref{Fig5} is summarizing our current efforts for making more complex and even more interesting devices.

Fig.~\ref{Fig5}a) shows an SEM micrograph of a multi-terminal suspended graphene flake with side contacts and pre-defined bottom gates, which is designed to explore the effect of electron guiding by gate defined potential \cite{CMguiding}. We have found that these devices can survive current annealing between the major contacts, and the graphene can also be cleaned around the side-contacts. Figure~\ref{Fig5}b) shows another multi terminal device. Here we have defined six contacts (1-6) to the graphene, four contacts at the edge, and finger-injectors at the middle of the flake. The finger injector (2,5) is separated from graphene by an aluminum oxide layer and only touches graphene in a small region. The structure is complemented with a bottom gate, shown in yellow. These and similar structures will allow to conduct electron-optics experiments in ultra-clean, ballistic graphene.
Figure~\ref{Fig5}c) presents an SEM micrograph of a suspended graphene flake, where a hole has been realized prior to suspension by oxygen plasma etching to realize a suspended Aharonov-Bohm ring. This shows that the graphene can be pre-shaped before suspension without the apparent collapse of the flake after suspension. However the yield of successful current annealing is decreasing. The results of these measurements will be published elsewhere.

\section{Conclusion}

In conclusion, we have shown that the method, based on LOR resist as a sacrificial layer to suspend graphene, can be supplemented with local gating to realize complex structures.  As the contacts are also suspended the cleaning of the graphene flake with current annealing can be performed with a high yield, and ultra-high mobilities can be obtained. The graphene can be supplemented with bottom and top-gates to form p-n junctions, which are the building blocks of more complex designs. The ballistic nature of the devices was demonstrated with Fabry-Perot interferences in micron-sized samples. Finally, further possibilities of the technique, such as side contacts and pre-shaping of the flake were shown.

\section{Acknowledgements}

This work was financed by the ESF programme Eurographene, the Swiss NSF, the EU FP7 project SE2ND, the ERC Advanced Investigator Grant QUEST, the ERC 258789, the Swiss NCCR Nano QSIT, and Graphene Flagship. We are grateful to Matthias Braeuninger, Andreas Baumgartner, Szabolcs Csonka, Ming-Hao Liu and Klaus Richter for fruitful discussions. We thank Alexander van der Torren and Dominik Bischoff for helping us to develop the graphene transfer method.

\bibliographystyle{elsarticle-num}
\bibliography{carbon}

\end{document}